# Reply to the comment on "Radiation forces and torque on a rigid elliptical cylinder in acoustical plane progressive and (quasi)standing waves with arbitrary incidence" [Phys. Fluids 28, 077104 (2016)]

F.G. Mitri*

(*Email: F.G.Mitri@ieee.org)

ABSTRACT

The aim of this communication is to correct inaccurate statements presented in a Commentary on the paper titled: "Radiation forces and torque on a rigid elliptical cylinder in acoustical plane progressive and (quasi)standing waves with arbitrary incidence" [Phys. Fluids **28**, 077104 (2016)].

The aim of this Reply is to correct inaccurate statements presented in a Commentary on the paper titled: "Radiation forces and torque on a rigid elliptical cylinder in acoustical plane progressive and (quasi)standing waves with arbitrary incidence" [Phys. Fluids 28, 077104 (2016)] (i.e., Ref.[1]). For the reader's convenience, it would be important to note the following points:

1- The peer-reviewed articles M1 [1], M2 [2], and M3 [3] deal, respectively, with *exact* solutions using the partial-wave series expansion (PWSE) method in cylindrical coordinates for **i)** the axial and transverse acoustic radiation forces and spin torque of progressive, standing and quasi-standing waves at any arbitrary angle of incidence $\alpha$ [1], **ii)** the axial acoustic radiation force of progressive, standing and quasi-standing waves at normal incidence (i.e. $\alpha = 0°$) [2], and **iii)** the acoustic scattering and its close connection with the radiation force and the scattering energy efficiency in the context of elliptical cylinders at normal incidence in plane progressive waves [3]. An extension to the case of the scattering of *finite acoustical beams* (in 2D) with any arbitrary angle of incidence on elliptical cylinders[4] has been also performed, but was not acknowledged in the Commentary. The intent of the works[1-4] was not to provide a review of all applications, rather to rigorously investigate in systematic studies all these important topics and phenomena from the standpoint of acoustic scattering, radiation force and torque theories using an *exact* analytical formalism without any approximations, while acknowledging some applications reviewed in Ref. [5].

2- Concerning the translational acoustic radiation forces per-unit-length on elliptical cylinders, it is important to emphasize that:

   a. The analytical expressions for the radiation force function of plane standing waves mentioned in Refs.[9-11] of the Commentary (which were extended for the cases of quasi-standing plane waves[6], cylindrically diverging waves[7] and later generalized to axisymmetric fields[8, 9] – but were not acknowledged in the Commentary), are *only* applicable for an angle of incidence $\alpha = n\pi/2$, with $n = 0, 1, 2$ …. at normal incidence. This case is known also as the axial/longitudinal case, such that there is no transverse force component. Notice, however, that when $\alpha$ deviates from $n\pi/2$ for an elliptical cylinder, the analytical expressions for the axial/longitudinal radiation force function of plane standing waves mentioned in Refs.[9-11] of the Commentary become inapplicable. Therefore, there is an essential need to develop a complete analytical theory and an improved rigorous formalism (not restricted to the axial/longitudinal case) applicable to elliptical cylinders, taking into consideration the character of the incident field, encompassing plane progressive, standing or quasi-standing waves at an arbitrary angle of incidence $\alpha \neq n\pi/2$. This progress has been developed and presented for the first time in M1 [1], for both the axial and transverse radiation force components as well as the acoustic radiation torque on elliptical cylinders without any approximations.

b. The results of Refs.[9-11] of the Commentary are only approximate, as they were restricted to an ellipse with very small relative deformation ($\varepsilon \ll 1$). Beyond that limit, the shape perturbation method used therein becomes inaccurate to compute correctly the radiation force (and torque) components, as the aspect ratio $b/a > 1$ or $< 1$, respectively. Though some discussions have been given therein for the elliptical cylinder case, the obtained scattering coefficients (which constitute the crux to compute the force components) are only *approximate*. In addition, Ref.[11] of the Commentary (which is a non-peer reviewed 4-pages conference proceeding) does not provide any plot or illustration of the results, rather it displays some radiation force approximations of very limited use, only applicable in the low-frequency domain (i.e., where the major or minor axes of the elliptical cylinder are much smaller than the illuminating wavelength) and for very small relative deformations $\varepsilon \ll 1$. These approximations are not applicable for the elliptical cylinder examples[1-4] examined using the multipole expansion method for larger aspect ratios. Fortunately, the generalized analysis M1 [1] considers an elliptical cylinder of any arbitrary aspect ratio $a/b$, by means of the rigorous mathematical formalism based on the *exact* partial-wave series expansion method in cylindrical coordinates, without restriction to the low-frequency (Rayleigh) limit. Therefore, the analytical formalism and results presented in M1 [1] (as well as in M2 [2] and M3 [3]) are generalized solutions obtained *without any approximations*.

c. Regarding the earlier work of Hasegawa[10] for a sphere (mentioned in the Commentary), which considered the case of plane quasi-standing waves as well as standing and progressive waves, it must be emphasized that the formalism (devoted initially for a sphere[10]) is *not* suitable by any means nor applicable to the circular/elliptical *cylindrical* geometry (in 2D). Fortunately, rigorous analyses for plane quasi-standing waves[6] and axisymmetric fields for cylinders were developed[8, 9], but were not acknowledged in the Commentary. It must be emphasized here that Eqs.(15) and (16) of M1 [1] constitute the most general expressions for the axial and transverse radiation force function components for any 2D object of arbitrary shape in plane waves of any type (i.e., progressive, standing or quasi-standing).

3- Concerning the radiation torque computations for rigid elliptical cylinders, the rigorous analysis M1 [1] does not provide approximations, rather *exact* calculations for the low-frequency limit (i.e., $kb \ll 1$), which have been adequately displayed in Figs. 4-6, in the third column, in the range $0 < kb \leq 5$, for all the chosen aspect ratios $a/b$ and reflection coefficient $R$ values. The case where $kb \ll 1$ is included in the plots, in contrast to what has been mentioned in the Commentary. Furthermore, the justification for evaluating the acoustic radiation torque using far-field scattering was originally given by Maidanik[11]. Notice that the torque exerted on a viscoelastic circular cylinder placed arbitrarily in a focused quasi-Gaussian beam has been also computed[12], but this was not acknowledged in the Commentary either.

4- For computational studies of acoustic radiation forces on cylinders as well as the scattering, absorption and extinction energy efficiencies contemporaneous with M1-M3 [1-3], the reader is referred to the recent analysis[13].

In conclusion, the thorough and rigorous analyses presented in M1-M3 [1-3] contain new and original results not presented previously in the scientific literature, nor in any other form of publication. The points discussed in this Reply placing the works M1-M3[1-3] in the proper context, should be helpful to the scientific community interested in the topic of acoustic radiation forces and torque on irregularly-shaped objects in 2D.